\documentclass[preprint]{aastex63}

\accepted{August 5 2021}
\submitjournal{PSJ}

\shorttitle{Tidal obliquity variations with \texttt{SMERCURY-T}}
\shortauthors{Kreyche, Barnes, Quarles, $\&$ Chambers}
\graphicspath{{./}{figures/}}

\usepackage[caption=false]{subfig}
\usepackage{bm}

\begin{document}

\title{Exploring tidal obliquity variations with \texttt{SMERCURY-T}}

\author[0000-0002-7274-758X]{Steven M. Kreyche}
\affiliation{Department of Physics, University of Idaho \\
Moscow, Idaho, USA}
\email{stevenkreyche@gmail.com}

\author[0000-0002-7755-3530]{Jason W. Barnes}
\affiliation{Department of Physics, University of Idaho \\
Moscow, Idaho, USA}

\author[0000-0002-9644-8330]{Billy L. Quarles}
\affiliation{Center for Relativistic Astrophysics, School of Physics, Georgia Institute of Technology  \\
Atlanta, Georgia, USA}

\author{John E. Chambers}
\affiliation{Department of Terrestrial Magnetism, Carnegie Institution of Washington\\
Washington, DC, USA}



\begin{abstract}

We introduce our new code, \texttt{SMERCURY-T}, which is based on existing codes \texttt{SMERCURY} \citep{Lissauer_2012} and \texttt{Mercury-T} \citep{Bolmont_2015}. The result is a mixed-variable symplectic N-body integrator that can compute the orbital and spin evolution of a planet within a multi-planet system under the influence of tidal spin torques from its star. We validate our implementation by comparing our experimental results to that of a secular model. As we demonstrate in a series of experiments, \texttt{SMERCURY-T} allows for the study of secular spin-orbit resonance crossings and captures for planets within complex multi-planet systems. These processes can drive a planet's spin state to evolve along vastly different pathways on its road toward tidal equilibrium, as tidal spin torques dampen the planet's spin rate and evolve its obliquity. Additionally, we show the results of a scenario that exemplifies the crossing of a chaotic region that exists as the overlap of two spin-orbit resonances. The test planet experiences violent and chaotic swings in its obliquity until its eventual escape from resonance as it tidally evolves. All of these processes are and have been important over the obliquity evolution of many bodies within the Solar System and beyond, and have implications for planetary climate and habitability. \texttt{SMERCURY-T} is a powerful and versatile tool that allows for further study of these phenomena.

\end{abstract}

\keywords{Exoplanet dynamics - Astrobiology - Habitable planets - Computational methods}


\section{Introduction} \label{sec:introduction}

Planetary obliquity is a fundamental parameter that factors into the assessment of a planet's habitability. The obliquity of a planet, or axial tilt, is the angle between the planet's spin axis and orbit normal. Together, the value of obliquity and its evolution over time determines the nature of the planet's seasons \citep{Williams_1997, Williams_2003, Kilic_2017, Kane_2017, Kang_2019}, atmospheric and oceanic processes \citep{Kilic_2018, Guendelman_2019, Dong_2019, Olson_2020}, and ultimately its climatic stability \citep{Milankovic_1998, Armstrong_2014, Spiegl_2015, Deitrick_2018b, Colose_2019}. Various studies investigated the obliquity evolution of bodies in the Solar System as well as exoplanets for purposes such as gaining insights into their dynamical history, making predictions for the future, or even judging their potential for habitability. 

Many processes can contribute to alter a planet's obliquity over time. Constant torques exerted on a planet's equatorial bulge by its planetary neighbors in addition to the regular precessional motion of the planet's orbit plane set a range of benign obliquity variations. On the other hand, a planet can experience large obliquity variations of many tens of degrees if it enters into a secular spin-orbit resonance. Such a resonance occurs when a planet's spin-axis precession becomes commensurate with a driving eigenfrequency of its orbital precession \citep{Colombo_1966, Peale_1969, Henrard_1987}. The resultant obliquity variations of this resonance depend on the amplitude of the orbital mode in addition to the planet's proximity to the resonance center. Provided that a planet resides within a spin-orbit resonance, its resonant angle, $\sigma = \psi + \phi$, will typically librate about 0 (the stable Cassini States 1 and 2 or the unstable Cassini State 4) or 180 degrees (Cassini State 2, stable), where $\psi$ is the planet's spin precession angle that defines the azimuthal orientation of the planet's spin vector and $\phi$ is the phase angle of the orbital precession mode \citep{Deitrick_2018a, Saillenfest_2019, Su_2020}. Furthermore, in the case that the widths of multiple spin-orbit resonances overlap, a planet's spin axis can be subject to chaotic evolution in which its obliquity swings across a wide range of values. Mars is the poster child for this chaotic behavior, as its obliquity swings from $\sim 0-60^\circ$ \citep{Laskar_1993b, Touma_1993, Laskar_2004b, Li_2014}. The Earth would suffer a similar fate if not for the Moon's help in speeding up the Earth's spin-axis precession enough so that the Earth peacefully resides outside of the chaotic region \citep{Laskar_1993a, Laskar_2004a, Brasser_2011}.

Considering the working parts that contribute to this resonance, rhythmic gravitational tugs from neighboring planets drive orbital precession while torques from the planet's star and moons fuel its spin precession. The rate of a planet's spin precession obeys

\begin{equation}
    \dot{\psi} = \frac{\alpha \cos{\epsilon} }{(1-e^2)^{\frac{3}{2}}}
    \label{eqn:prec_freq}
\end{equation}

\noindent where $\epsilon$ is the planet's obliquity, and $e$ is its orbital eccentricity. In the case of a moonless planet, the precession constant, $\alpha$, goes as

\begin{equation}
    \alpha \approx \frac{3n^2}{2\omega_p} \frac{J_2}{\overline{C}}
    \label{eqn:prec_const}
\end{equation}

\noindent Here $n$ is the planet's mean motion, $\omega_p$ its spin rate, $J_2$ its oblateness coefficient, and $\overline{C}$ is its normalized polar moment of inertia (this expression is approximate, based on the assumption that the planet's dynamical ellipticity is roughly proportional to ${\omega_p}^2$) \citep{Laskar_1993b, Li_2014, Quarles_2020}. 

The status of a planet's presence within a spin-orbit resonance can change with time. For instance, although the Earth rests easy now, it may have crossed a high-order resonance just a few million years ago \citep{Laskar_1993a, Laskar_2004a}. Change can be brought on by a variety of processes, such as those that shift the location and amplitudes of potentially resonant orbital frequencies. Planetary orbital migration mechanisms are one such way to achieve these shifts \citep{Brasser_2015, Vokrouhlick__2015, Millholland_2019a}. In a similar vein, the works of \citet{Saillenfest_2020}, \citet{Saillenfest_2021a}, and \citet{Saillenfest_2021b} studied the past and future influence that satellite migration may have had on the obliquity evolution of Jupiter and Saturn. \citet{Millholland_2019b} and \citet{Su_2020} studied another mechanism in which a shrinking protoplanetary disk could cause sweeping spin-orbit resonances to influence the obliquity distribution of newly formed protoplanets. Another study applied this concept to Uranus's circumplanetary disk to attempt to infer its dynamical history \citep{Rogoszinski_2020}. 

Alternatively, a planet can be driven into or out of a spin-orbit resonance by processes that alter its spin precession rate (Equation \ref{eqn:prec_freq}). Tidal spin torques are one mechanism that can dynamically shape the fate of a planet's spin state over time. The tidal friction generated from a system's star pulling on the equatorial bulges of a planet will slowly drive it toward tidal equilibrium, in which its obliquity will typically increase followed by a gradual decline to a small values, while its spin rate undergoes rotation braking until its spin period becomes synchronous with its orbital period (or pseudo-synchronous for eccentric orbits). This journey amounts to a constant decline in a planet's precession constant and consequently its spin-precession rate over time, allowing for opportunities to encounter spin-orbit resonances as time goes on. Tidal spin torques can drive a planet's spin axis to traverse parameter space and undergo scenarios such as spin-orbit resonance crossings, captures, or even traverses of chaotic resonance regions. Since the magnitude of the tidal force decays roughly as $r^{-6}$ with distance, this process is likely especially important for planets in close proximity to their star. The Earth will likely undergo chaotic obliquity variations itself in a few billion years after entering the chaotic zone thanks to its decaying rotation rate \citep{Neron_de_Surgy_1997}. 

We introduce \texttt{SMERCURY-T} \citep{Kreyche_2021}, a joining of the existing code \texttt{SMERCURY} \citep{Lissauer_2012} with elements from \texttt{Mercury-T} \citep{Bolmont_2016} so that we can investigate these dynamic tidal spin phenomena and build a more complete picture of planetary spin dynamics. While \texttt{SMERCURY} utilizes its implementation of the spin Hamiltonian from \citet{Touma_1994} to accurately compute the orbital and rotational dynamics of a planetary system due to N-body interactions, it is not equipped to compute tidal evolution. On the other hand, \texttt{Mercury-T} is capable of computing the tidal evolution of a system but limits the study planetary spin evolution by neglecting planet-planet cross terms in its spin calculations. In addition, \texttt{Mercury-T} lacks a scheme to efficiently explore parameter space like the ``ghost planet" scheme featured in the \texttt{SMERCURY-T} code (discussed in Section \ref{sec:old}). Similar open-source codes such as \texttt{Posidonius} \citep{Blanco_2017} and \texttt{VPLanet} \citep{Barnes_2020} are also guilty of these charges. We bridge this gap by branching \texttt{SMERCURY-T} from \texttt{SMERCURY} to include modifications that allow for the complete simultaneous computation of the orbital and spin evolution of a planet while under the effects of stellar tidal spin torques.

In this article, we begin with Section \ref{sec:methods} in which we describe the \texttt{SMERCURY-T} code and discuss our implementation of its new modules. Following that, in Section \ref{sec:experiments} we perform several experiments that demonstrate the phenomena of spin-orbit resonance crossings, captures, and chaotic crossings which express the versatility of \texttt{SMERCURY-T} as a tool. We conclude in Section \ref{sec:conclusion}. 

\section{Methods} \label{sec:methods}

\subsection{The Base Code} \label{sec:old}
We present our new code, \texttt{SMERCURY-T}, as a modified version of its predecessor, \texttt{SMERCURY} \citep{Lissauer_2012, Barnes_2016, Quarles_2020, Kreyche_2020}. \texttt{SMERCURY} is a mixed-variable symplectic N-body integrator which is itself a modified spin-tracking version of the original \texttt{MERCURY} package \citep{Chambers_1999}. Just like the code that \texttt{SMERCURY-T} derives from, it is capable of computing the orbital evolution of multiplanet systems while tracking the spin state of one planet and its ``ghosts". The ghost planet scheme entails the simultaneous simulation of massless copies of the spin-tracked planet, in which the user assigns a unique initial spin state to each ghost planet which will evolve while its orbital evolution mimics that of the original planet. This scheme makes \texttt{SMERCURY-T} advantageous over similar codes by allowing for efficient exploration of parameter space, however its present implementation restricts the user to only being able to simulate the spin evolution of one system's planet at a time. \texttt{SMERCURY-T} is publicly available \citep{Kreyche_2021}.

\texttt{SMERCURY-T} treats the spin computation by considering the spin-tracked planet to be an axisymmetric rigid body. Gravitational torques from the planet's star and neighboring planets exerted on the spin-tracked planet's equatorial bulge modify its obliquity. When preparing a simulation, we express the planet's spin rate by assigning it a value for its zonal gravity coefficient $J_2$ according to the Darwin-Radau relation \citep{Hubbard_1984, Murray_2000}. This takes the form

\begin{equation} \label{eqn:J_2}
    J_2 = \frac{\omega_p^2{R}^3}{3Gm_p} \left (\frac{5}{D } - 1 \right )
\end{equation}

\noindent with $\omega_p$, $R$, and $m_p$ as the planet's spin rate, mass, and equatorial radius, respectively. Here $G$ is the universal gravitational constant and the placeholder $D$ is short for

\begin{equation}
    D = \frac{25}{4} \left (\frac{3}{2}\overline{C}-1 \right)^2 + 1
\end{equation}

\noindent where $\overline{C}$ is the planet's normalized polar moment of inertia. We calculate $R$ according to

\begin{equation} \label{eqn:R}
    R = \left [\frac{G m_p D}{{\omega_p}^2} \left (\frac{2-\sqrt{4-\frac{30{\omega_p}^2}{G D \pi \rho}}}{10} \right) \right]^\frac{1}{3} 
\end{equation}

\noindent with $\rho$ as the average density of the planet. We refer to \citet{Lissauer_2012} for a more thorough description of this technique.

\subsection{Code Modifications} \label{sec:new}
In this section, we describe the modifications that we include within the \texttt{SMERCURY-T} package based on the alterations that we made to the base \texttt{SMERCURY} code. Here we include the details and implementation of the tidal spin torque module and then discuss the results of a test experiment to verify its accuracy. We also discuss our additional inclusion of a general relativity module.

\subsubsection{Tidal Spin Torque Module} \label{sec:obl_tides_module}
The most significant inclusion to the \texttt{SMERCURY-T} code is the tidal spin torque module. We follow the methods of \citet{Bolmont_2015} and adopt the constant time lag tidal formulation expressed by \citet{Leconte_2010} and others \citep{Mignard_1979, Hut_1981, Eggleton_1998}, which is valid for arbitrary values of obliquity, spin, and orbital eccentricity. Contrary to the rigid body considerations that we employ for the N-body torque computations, this model assumes that the body is made of a weakly viscous fluid \citep{Alexander_1973}. In the future, SMERCURY-T's modular design makes it simple to implement and explore alternative tidal models. Our approach differs from the aforementioned works in that we neglect orbital tides and consider only the effects of tidal spin torque on the spin-tracked planet from its star. This choice saves valuable computation time, while still allowing for sufficient study of the behavior of planetary spin evolution in most planetary systems except the most compact ones. Readers can refer to Table 2 of \cite{Mardling_2004}, or can perform quick calculations with handy equations such as those found in \citet{Rodriguez_2010} to estimate the importance of orbital tides in their desired system. We justify our study of the systems that we consider in Section \ref{sec:experiments} by estimating the semi-major axis damping timescale ($a/\dot{a}$) and eccentricity damping timescale ($e/\dot{e}$) to be practically infinite for the Earth-mass planet under study compared to our integration time of 4 Gyr.  Another consequence of our methodology is that it is technically in violation with the law of conservation of angular momentum, since the spin-tracked planet's spin angular momentum decays while the system gets nothing back in return. The choice to neglect the effects of this feedback is acceptable for the study of planetary systems since the total angular momentum of the system will always vastly dwarf the the spin-tracked planet's spin angular momentum.

The task to modify the spin routine of \texttt{SMERCURY-T} to include tidal spin torques is a relatively straight-forward one. Since \texttt{SMERCURY-T} already integrates the summation of all of the torques felt by the spin-tracket planet's rotational bulge at each time step, all that remains is calculating the additional tidal torque within the routine. From \citet{Bolmont_2015}, the non-averaged tidal torque felt by a planet from its star goes as 

\begin{equation} \label{eqn:tidal_torque}
    \mathbf{N_T} = 3G \frac{{m_s}^2 {R}^5}{r^7} k_2 \tau (r \bm{\omega}_p - (\mathbf{r} \cdot \bm{\omega}_p)\mathbf{e}_r - \mathbf{e}_r \times \mathbf{v}) 
\end{equation}

\noindent where $\mathbf{r}$ is the radial vector between the planet at its star, $\bm{\omega}_p$ is the planet's spin vector, $\mathbf{e_r}$ is the radial unit vector, $\mathbf{v}$ is the planet's velocity vector, $k_2$ is its potential love number of degree 2, $\tau$ is the constant time lag, $m_s$ is the mass of the star, and $G$ is the universal gravitational constant. This calculation treats the star as a point mass while the planet is made up of a reduced central point mass with two point mass bulges. Figure 1 from \citet{Bolmont_2015} is a useful diagram that demonstrates some of these elements. Converting this result to heliocentric coordinates allows for a final expression that conforms with the requirements of \texttt{SMERCURY-T}'s spin algorithm. Taking the planet's spin angular momentum as $\mathbf{S}$, the adjusted tidal torque is

\begin{equation}
    \frac{d}{dt} \mathbf{S} = -\frac{m_s}{m_s + m_p} \mathbf{N_T}
\end{equation}

As the spin-tracked planet's spin rate damps over time due to the inclusion of the tidal torque, it is necessary to update the planet's $J_2$ and equatorial radius at each time step to represent the reduction of its equatorial bulge. This is important not only due to the dependence of the strength of the tidal torque on $R$, but even more so due to the sensitive nature of the N-body spin dynamics computations to the value of $J_2$ as the planet's spin state evolves. We develop a scheme to update these parameters and provide the details in Appendix \ref{sec:appendix_a}. This scheme involves instructing \texttt{SMERCURY-T} to use the value of the planet's spin angular momentum to solve a cubic equation to ultimately determine new values for $R$ and $J_2$ using Equations \ref{eqn:R_solve} and \ref{eqn:J_2}. 

One possible concern with the inclusion of the tidal spin torque module is that there may be scenarios in which the magnitude of the computed tidal torque is so small that it under flows machine precision, causing the planet's tidal evolution to cease entirely. We solve this issue by implementing a ``tidal tolerance" input parameter that the user can specify prior to running a simulation. This parameter is a fractional value that determines a certain threshold after being multiplied by the total spin angular momentum at the current time step. If the tidal torque components exceed this threshold, then they are integrated and added to the spin vector. However, if they fall below the threshold then \texttt{SMERCURY-T} saves their values and adds them to a running total to be checked the next time around. A typical value for the tidal tolerance parameter that we use in this article is $10^{-12}$.

\subsection{Tidal Spin Torque Module Verification} \label{sec:verification}
We verify the implementation of the \texttt{SMERCURY-T} tidal spin torque module by simulating and comparing the results of a test system's tidal evolution to that of a secular code based on the framework of \citet{Leconte_2010}. The secular code is capable of solving the averaged equations of one planet's orbital and rotational tidal evolution. However, since \texttt{SMERCURY-T} only accounts for the effects of tidal spin torques, for a fair comparison we hold the orbital elements constant and evolve the secular code solely with the spin equations. We implement the same scheme from Section \ref{sec:obl_tides_module}, detailed in Appendix \ref{sec:appendix_a}, into the secular code to similarly update the planet's equatorial radius at each time step.

The test system that we use here is a one-planet system that consists of the Sun and an Earth-mass planet with an semi-major axis of 0.2 AU and an orbital eccentricity of 0.3. This choice ensures a short integration time and demonstrates the capability of \texttt{SMERCURY-T} to handle large eccentricities. For both simulations, we assign this planet with a polar moment of inertia, average density, initial spin period, initial equatorial radius, potential love number of degree 2, constant time lag, and initial obliquity according to Table \ref{tab:physical}. We integrate the \texttt{SMERCURY-T} simulation with a time step of 1.6 days ($\sim$ 5\% of the inner-most planet's orbital period, our usual rule of thumb) and sample at an interval of 1000 years. The time step for the averaged equations of the secular code does not need to be as small, so we proceed with 1000 year increments. We run both simulations for 10 Myr which is more than enough time for the planet to reach tidal equilibrium.

\begin{deluxetable}{c c c c c c c}[!ht]
\tabletypesize{\footnotesize}
\tablecaption{\\ Planet Physical Parameters \label{tab:physical}}
\tablehead{\colhead{$\overline{C}$} & \colhead{$\rho$ [g/$cm^3$]} & \colhead{$P_0$ [hours]} & \colhead{$R_0$ [km]} & \colhead{$k_2$} & \colhead{$\tau$ [sec]} & \colhead{$\epsilon_0$ [deg]}} 
\startdata
0.3296108 & 5.5136 & 24 & 6378 & 0.305 & 698 & 45 \\
\enddata
\tablecomments{Physical values for the test planet that we describe in Section \ref{sec:verification}. Here we list the planet's moment of inertia ($\overline{C}$), average density ($\rho$), initial spin period ($P_0$), initial equatorial radius ($R_0$), potential love number of degree 2 ($k_2$), constant time lag ($\tau$), and initial obliquity ($\epsilon_0$).}
\end{deluxetable}

\begin{figure}[!ht]
\centering
\includegraphics[width=\textwidth]{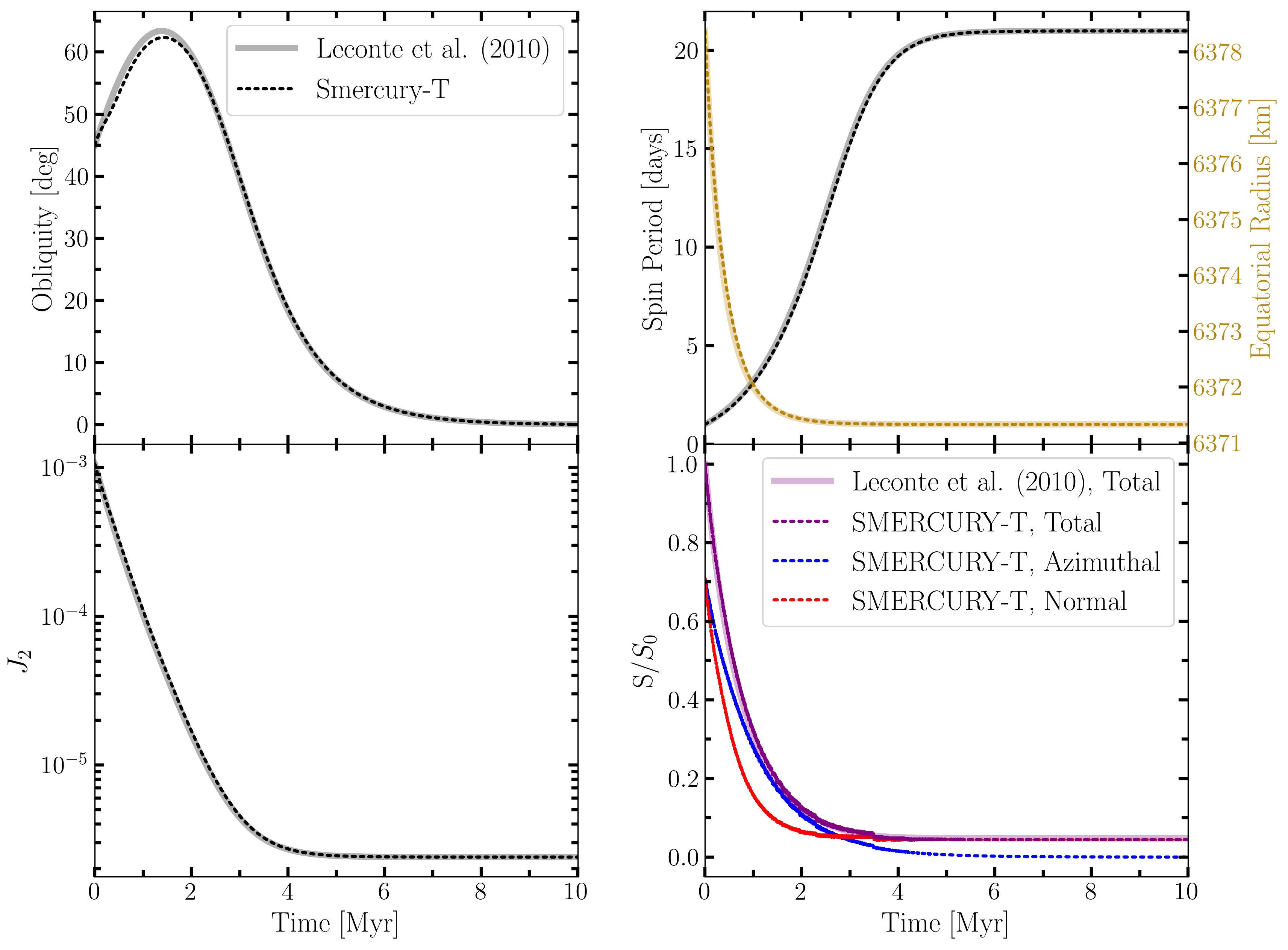}
\caption{A comparison between \texttt{SMERCURY-T} (dashed lines) and the secular tidal model from \citet{Leconte_2010} (solid lines) for the spin evolution of an Earth-mass planet orbiting at 0.2 AU with an eccentricity of 0.3. Tidal spin torques cause the planet's obliquity (top left), spin period and equatorial radius (top right), zonal gravity coefficient, $J_2$ (bottom left), and spin angular momentum, $S/S_0$ (expressed as a fraction of the total initial spin angular momentum, bottom right), to evolve over the course of each 10 Myr simulation. For the \texttt{SMERCURY-T} result, we also show the breakdown of the planet's total spin angular momentum into its azimuthal (dashed blue line) and normal components (dashed red line).}
\label{fig:compare}
\end{figure}

Looking to Figure \ref{fig:compare}, we show a comparison between the results of the two codes for the spin evolution of the Earth-mass planet due to the effects of tidal spin torques. Examining the evolution of the planet's obliquity, spin period, equatorial radius, $J_2$ (initially calculated with Equation \ref{eqn:J_2}), and spin angular momentum, we show that \texttt{SMERCURY-T} is in good agreement with the secular code. Although there is a slight discrepancy between the extent of the planet's obliquity values near the peaks of the plot in Figure \ref{fig:compare}'s upper left panel, each parameter ultimately tracks and converges to the same end value with little difference between the two codes. In summary, we see that tidal spin torques cause the planet to initially experience an increase in its obliquity until its spin rate decays such that the altitudinal component of its spin angular momentum overtakes the azimuthal component, as shown in the upper and lower right panels of Figure \ref{fig:compare}. From there, its spin period continues to increase while its obliquity declines until it reaches tidal equilibrium when its obliquity is near zero and its spin period is that of its pseudo-synchronous period ($\sim$ 21 days here). The planet's equatorial radius shrinks by about 7 km while its $J_2$ reduces considerably.

\subsubsection{General Relativity Module} \label{sec:GR}
We also incorporate the general relativistic (GR) force due to the post-Newtonian potential into \texttt{SMERCURY-T} as an optional module following section 2.2 of \citet{Bolmont_2016}. The GR force not only plays an important role with regard to the stability of a planetary system, but we reason that it can also have an indirect influence on planetary obliquity evolution. This influence stems from the GR effect that causes the precession of periapsis of planetary orbits to speed up. Bearing this in mind, GR can indirectly influence planetary obliquity by shifting the potentially resonant orbital eigenfrequencies. While these effects are important to a degree in our own Solar System, they are likely especially important for compact exoplanet systems whose planets reside deep within their star's gravitational well. 
 
In the spirit of one famous confirmation of Einstein's general theory of relativity \citep{Einstein_1916}, we verify the implementation of our GR module by checking the contribution of GR to the precession of periapsis of our planetary neighbor, Mercury. We compare the results of our \texttt{SMERCURY-T} simulation to that of the measured rate from the \textit{MESSENGER} spacecraft reported by \citet{Park_2017}. We obtain an average value of $42.9790$ $''$/Julian century over the course of our 1 Myr simulation in which we used a system consisting only of the Sun and Mercury to isolate the effect of GR. Specifically, this value is the difference between Mercury's precession rate when the GR module was enabled versus another simulation when it was not. Here we set the integration time step to be 5\% of Mercury's orbital period and sampled at 100 year intervals. This value is in good agreement and lies within the error bars of the \citet{Park_2017} result.

\section{Experiments} \label{sec:experiments}
In this section, we apply the code modifications discussed in Section \ref{sec:new} to demonstrate the usefulness and versatility of \texttt{SMERCURY-T} as a tool to study planetary spin dynamics. Here we showcase examples of three different scenarios driven by the effects of tidal spin torques: a spin-orbit resonance crossing, a spin-orbit resonance capture, and a chaotic resonance crossing. We enable the general relativity and tidal spin torque modules (``tidal tolerance" set to $10^{-12}$) for these experiments with the exception of the initial frequency analysis runs detailed in Sections \ref{sec:crossing_initial} and \ref{sec:chaotic_initial}.

\subsection{Spin-orbit Resonance Crossing} \label{sec:crossing}

\subsubsection{Initial Conditions and Pre-analysis} \label{sec:crossing_initial}
We design an experiment to exemplify the scenario in which tidal spin torques drive a planet to undergo a spin-orbit resonance crossing as its increasing obliquity and spin period cause its precession constant to decline over time. We choose to use a simple toy system consisting of the Sun, an Earth-mass planet initially placed at 0.5 AU, and a Jupiter-mass planet initially placed at 2.5 AU. This choice allows us to study this process appropriately while minimizing the necessary integration time thanks to the stronger presence of tides closer to the Sun. We display the masses and orbital elements of these bodies in Table \ref{tab:orbital} listed under system E-J, which shows that the Earth-mass planet and Jupiter-mass planet share an initial mutual inclination of $1^\circ$.

\begin{deluxetable}{c | c c c c c c c c}[!ht]
\tabletypesize{\footnotesize}
\tablecaption{\\ System Orbital Parameters \label{tab:orbital}}
\tablehead{\colhead{System} & \colhead{Planet} & \colhead{$m_p$ [$M_{\odot}$]} & \colhead{$a$ [AU]} & \colhead{$e$} & \colhead{$I$ [deg]} & \colhead{$\omega$ [deg]} & \colhead{$\Omega$ [deg]} & \colhead{$M$ [deg]}} 
\startdata
E-J & Earth & 3.003 x $10^{-6}$ & 0.5 & 0 & 1 & 0 & 0 & 0 \\
  & Jupiter & 9.54244 x $10^{-4}$ & 2.5 & 0 & 0 & 0 & 0 & 0 \\
 \hline
E-E-J & Inner Earth & 3.003 x $10^{-6}$ & 0.5 & 0 & 5 & 0 & 0 & 0 \\
  & Outer Earth & 3.003 x $10^{-6}$ & 0.7 & 0 & 2.5 &  0 & 0 & 0 \\
  & Jupiter & 9.54244 x $10^{-4}$ & 3.5 & 0 & 0 & 0 & 0 & 0 \\
\enddata
\tablecomments{Initial values for the plantary mass ($m_p$), semimajor axis ($a$), eccentricity ($e$), inclination ($I$), argument of periapsis ($\omega$), longitude of ascending node ($\Omega$), and mean anomomly ($M$) for the Earthmoo-Jupiter (E-J)  and Earth-Earth-Jupiter (E-E-J) systems that we test in this article. We take the mass of Sun to be 1.98911$\times 10^{30}$ kg.}
\end{deluxetable}

We perform an initial simulation of the E-J system over the course of 10 Myr, sampling on an interval of 1000 years with a numerical time step of 6.46 days ($\sim$ 5\% of the Earth-mass planet's orbital period). For this part we are only interested in the system's orbital evolution, so we leave the GR module enabled but disable the tidal spin torque module in the interest of shaving off some computation time. Taking these results, we perform a frequency analysis routine by applying a Frequency Modified Fourier Transform \citep{Sidlichovsky_1996} to the Earth-mass planet's inclination and eccentricity vectors ([$\sin{\frac{I}{2}}\cos{\Omega}$, $\sin{\frac{I}{2}}\sin{\Omega}$] and [$e\cos{\varpi}$, $e\sin{\varpi}$] respectively). This process extracts the frequencies, amplitudes, and initial phases of the precessional modes that could enter resonance with the Earth-mass planet. First order resonances appear directly in the inclination series. We display these values in Table \ref{tab:freqs} listed under system E-J. This information allows us to compute both the location of a resonance center and its width enclosed by the separatrix as a function of a planet's precession constant and obliquity, following the procedure of \citet{Saillenfest_2019}. The simplicity of the E-J system results in the single dominant $\nu_1$ mode in the inclination series at $\approx -22.69$ $''$/yr. This frequency sets the target resonance for the Earth-mass planet in the experiments we describe in Section \ref{sec:crossing_results} and \ref{sec:capture_results}.

\begin{deluxetable}{c | c | ccc | ccc }[!ht]
\tabletypesize{\footnotesize}
\tablecaption{\\ Secular Orbital Frequencies \label{tab:freqs}}
\tablehead{\colhead{} & \colhead{} & \colhead{} & \colhead{Inclination Vector Series} & \colhead{} & \colhead{} & \colhead{Eccentricity Vector Series} & \colhead{} \\
\hline
\colhead{System} & \colhead{$j$} & \colhead{$\nu_j$ [$''$/yr]} & \colhead{$S_j \times 10^8$} & \colhead{${\phi^{(0)}_j}$ [deg]} & \colhead{$\mu_j$ [$''$/yr]} & \colhead{$E_j \times 10^8$} & \colhead{$\theta^{(0)}_j$ [deg]}  }
\startdata
E-J & 1 & -22.69 & 871406 & 359.94 & 22.89 & 804.86 & 179.80 \\ 
  & 2 & 0.00 & 1221 & 0.00 & 0.03 & 431 & 183.31 \\ 
  & 3 & -22.47 & 183 & 15.31 & -9.11 & 15 & 41.66 \\
  & 4 & -22.90 & 170 & 149.56 & -6.23 & 14 & 346.47 \\
  & 5 & -22.30 & 102 & 6.21 & -7.45 & 13 & 235.11 \\
\hline
E-E-J & 1 & -10.67 & 3736801 & 359.99 & 30.74 & 704 & 180.09 \\ 
  & 2 & -32.25 & 617977 & 359.84 & 0.03 & 617 & 180.22 \\ 
  & 3 & -3.97 & 8169 & 0.09 & 12.59 & 53 & 2.33 \\
  & 4 & -53.83 & 2096 & 179.43 & 9.15 & 14 & 191.46 \\
  & 5 & 10.91 & 786 & 358.52 & -98.64 & 8 & 77.37 \\
\enddata
\tablecomments{The results from our Frequency Modified Fourier Transform analysis of the inclination vector ([$\sin{\frac{I}{2}}\cos{\Omega}$,$\sin{\frac{I}{2}}\sin{\Omega}$]) and the eccentricity vector ([$e\cos{\varpi}$, $e\sin{\varpi}$]) of the Earth-mass planet in the Earth-Jupiter (E-J) system and the inner Earth-mass planet in the Earth-Earth-Jupiter (E-E-J) system, where $I$ is the orbital inclination, $\Omega$ is the longitude of ascending node, $e$ is the eccentricity, and $\varpi$ is the longitude of periapsis. We show the top five values of the frequencies ($\nu_j$), amplitudes ($S_j$), and phases (${\phi^{(0)}_j}$) of the inclination vector, as well as the top five values of the frequencies ($\mu_j$), amplitudes ($E_j$), and phases ($\theta^{(0)}_j$) of the eccentricity vector.}
\end{deluxetable}

Targeting the $\nu_1$ mode, we initialize the Earth-mass planet so that it will begin its tidal evolution just outside of the resonance and cross paths with the resonance center some time later. Therefore we assign it with an initial obliquity of $50^\circ$ and a precession constant of 60 $''$/yr (using Equation \ref{eqn:prec_const} with a spin period of $\approx 54.7$ hours and $J_2=2.03779 \times 10^{-4}$). These planetary characteristics are not unrealistic, and could be the aftermath of various primordial scenarios such as giant impacts. We set its initial precession angle to be $0^\circ$, while we use the same values for its average density, normalized polar moment of inertia, potential love number of degree 2, and constant time lag as found in Table \ref{tab:physical}. We run this simulation for 4 Gyr and sample every 10,000 years with the same 6.46 day time step.

\subsubsection{Results and Discussion} \label{sec:crossing_results}
\begin{figure}[!ht]
\centering
\includegraphics[width=\textwidth]{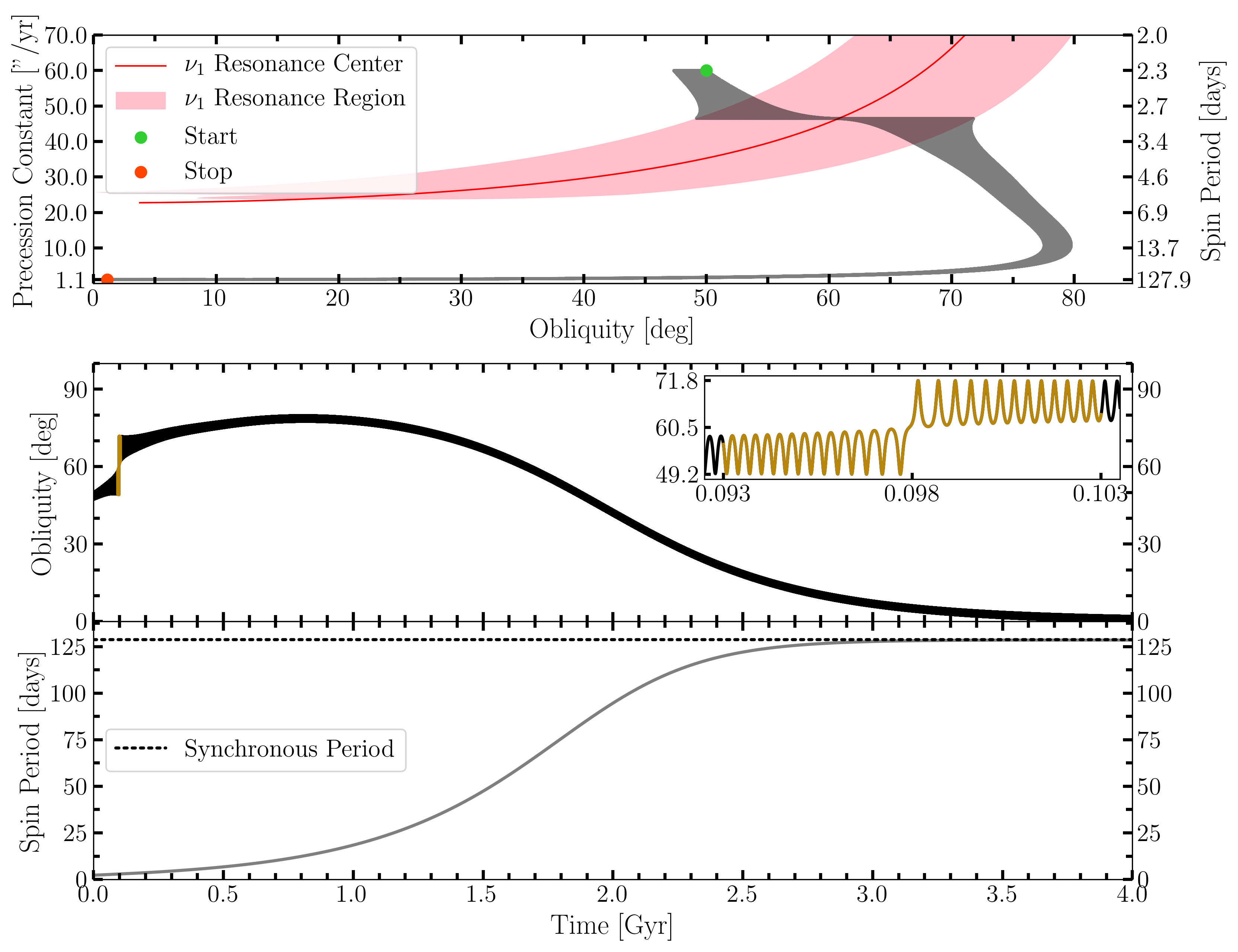}
\caption{We show the evolution of the Earth-mass planet's precession constant, obliquity, and rotation period over the course of 4 Gyr for the Earth-Jupiter system described in Table \ref{tab:orbital}. Here the planet's initial precession angle was $0^\circ$. In the upper panel the green and red dots indicate the start and stop points of the simulation while the black line traces out the spin evolution of the Earth-mass planet as tidal spin torques cause it to traverse through the pink-filled spin-orbit resonance region and across the $\nu_1$ resonance center (Cassini state 2, solid red line). The middle panel isolates the obliquity evolution and includes an inset that showcases the dramatic shift in obliquity around the time of the resonance crossing event (highlighted in gold). The lower panel shows the evolution of the spin period over time and marks the location of the planet's synchronous period (dashed black line).}
\label{fig:crossing_plot}
\end{figure}

\begin{figure}[!ht]
  \centering
  \includegraphics[width=0.75\textwidth]{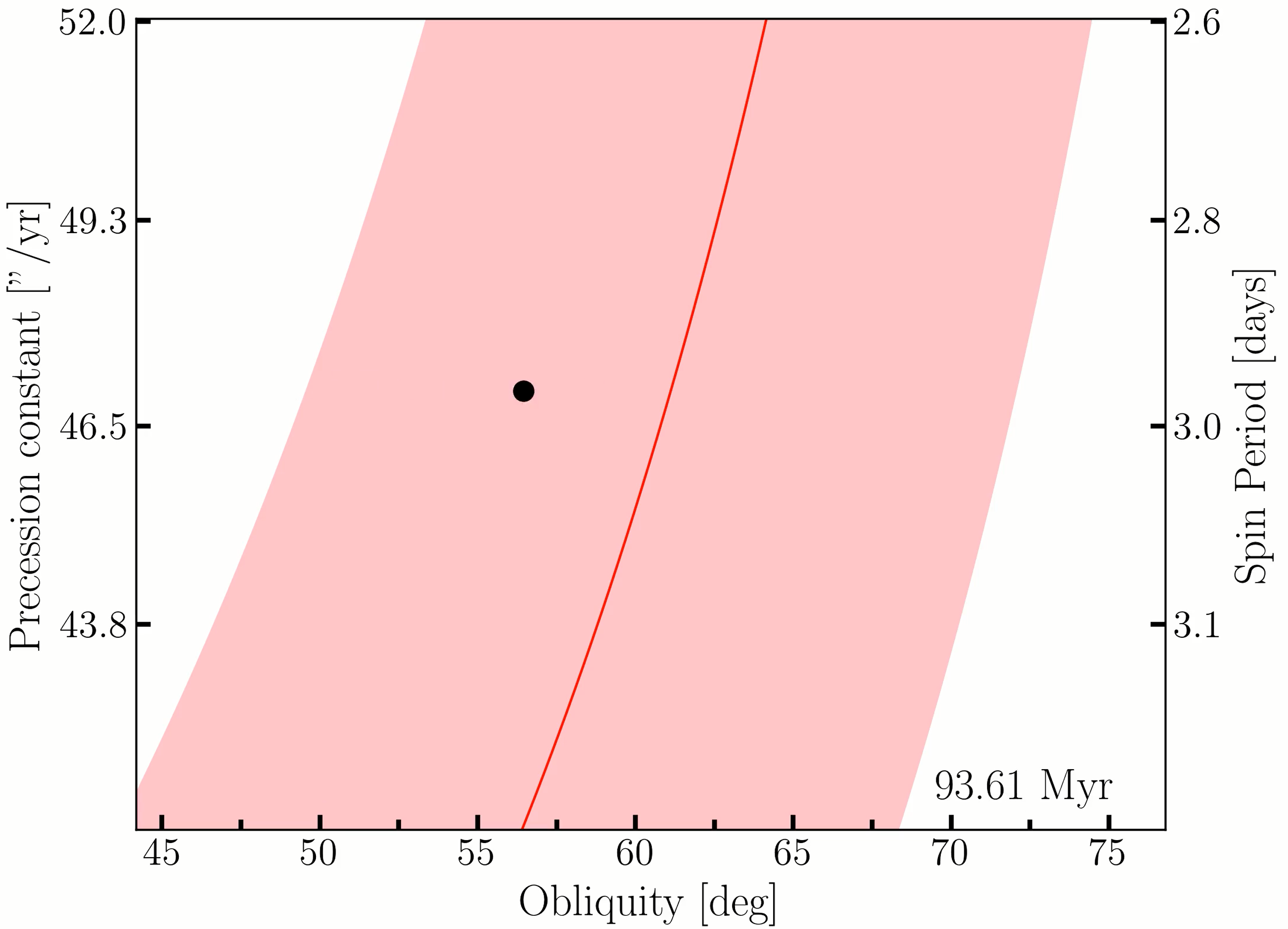}
  \caption{The real-time duration of this video is 30 seconds, during which we show a time lapse of the results from the top panel of Figure \ref{fig:crossing_plot}; refer to the caption of Figure \ref{fig:crossing_plot} for a description of the chart elements. Here we represent the Earth-mass planet's parameter position with a black dot and display an associated time stamp for each frame in the lower right corner. From the range 93-103 Myr into the simulation, tidal spin torques from the star cause the Earth-mass planet's precession constant to decline while it experiences strong obliquity variations as it traverses the resonance region. The planet's obliquity receives a prompt kick upon crossing the resonance center from the left to right side and then begins to move away from the resonance center afterwards. \\
  \url{https://github.com/SMKreyche/Kreyche-et-al.-2021-Videos}}
  \label{fig:crossing_frame}
\end{figure}

We explore the results of the spin-orbit resonance crossing experiment by looking to Figure \ref{fig:crossing_plot}. We see that the Earth-mass planet begins with an obliquity of $50^\circ$ and experiences benign obliquity variations of a few degrees that are primarily due to the precessional motion of its orbit. The planet's precession constant declines while tidal spin torques cause its obliquity to gradually increase as it approaches the resonance. Upon entering the resonance region, the planet's obliquity swings wildly in the range of $\sim 50-60^\circ$. Then, at about 98 Myr into the simulation, the planet's obliquity suddenly shifts ten degrees as it crosses the resonance center and swings in a new range of $\sim 60-70^\circ$. The planet then proceeds to exit the resonance and eventually converges towards tidal equilibrium with the behavior of Cassini State 1, as its obliquity draws near zero and its spin period approaches the synchronous period of $\sim 129$ days. The nature of these short term obliquity variations as the planet crossed through this resonance during its tidal evolution would likely be harmful toward its prospects for climatic stability and habitability. However, a true assessment of these prospects would require a more complete study that that would explore the application of a climate model. We include an accessory video that shows a time lapse of the crossing event in Figure \ref{fig:crossing_frame}.

\begin{figure}[!ht]
\centering
\includegraphics[width=\textwidth]{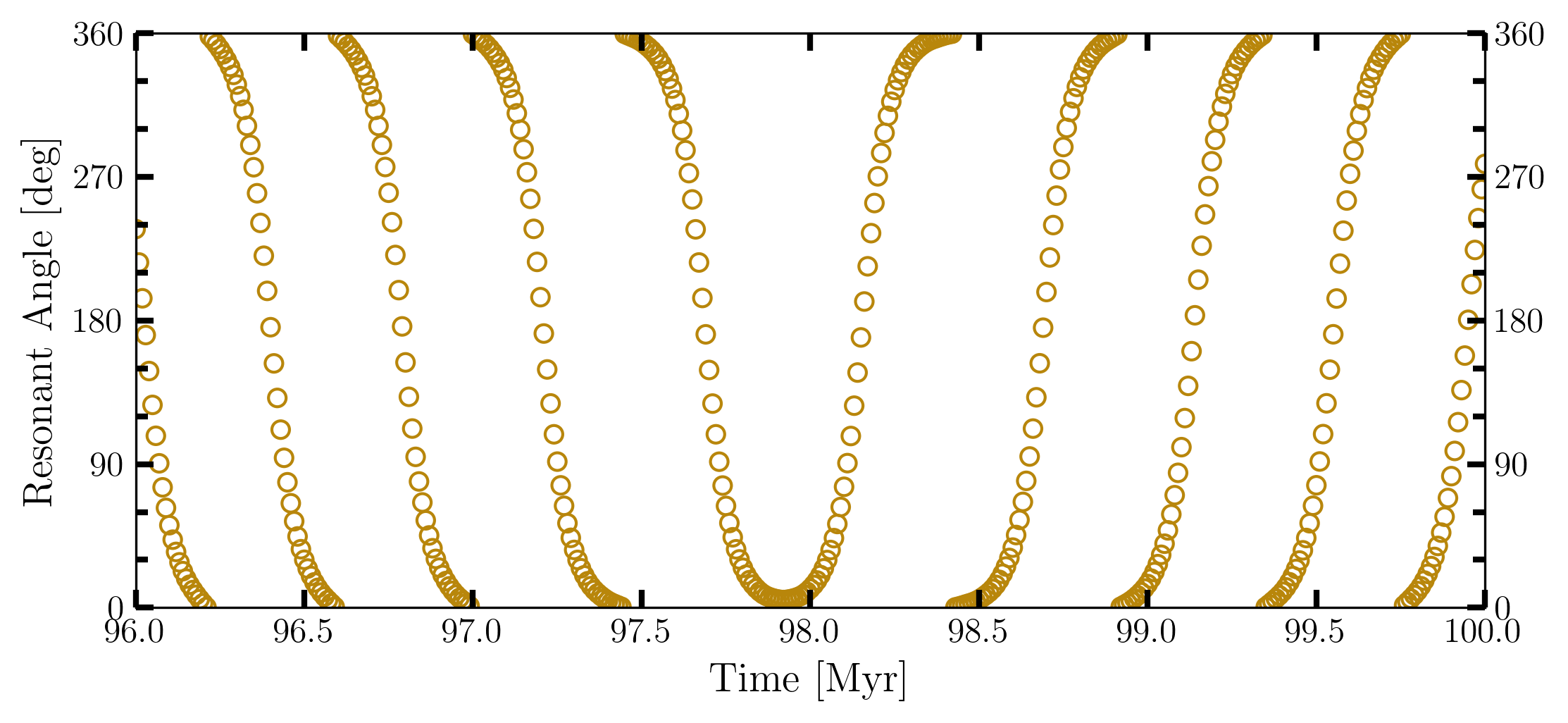}
\caption{We show the evolution of the Earth-mass planet's (from the Earth-Jupiter system described in Table \ref{tab:orbital}) resonant angle over time centered about the time that the planet crossed the $\nu_1$ resonance center. In this case, the planet's initial precession angle, $\psi$, was $0^\circ$. For the resonant angle we use $\sigma = \psi - \Omega$, with $\Omega$ as the planet's longitude of ascending node.} This angle is diagnostic of the presence of a planet in a spin-orbit resonance. Here the angle circulates with a brief inflection point at the time of the crossing event ($\sim 98$ Myr), which indicates that the planet was not captured into the resonance and instead crossed through its hyperbolic point \citep{Saillenfest_2021a}.
\label{fig:crossing_analysis}
\end{figure}

Further examination of the resonance crossing event leads us to reason that our choice of the planet's initial precession angle ensured that it crossed through the resonance's hyperbolic point \citep{Saillenfest_2021a}. In other words, this behavior resulted due to the chance timing and positioning of the planet's spin axis as its spin precession wound down while the planet crossed the resonance center. Looking to Figure \ref{fig:crossing_analysis}, we examine the evolution of the angle $\sigma = \psi - \Omega $ centered around the time of the resonance center crossing event at $\sim 98$ Myr, where $\psi$ is its spin precession angle and $\Omega$ is the planet's longitude of ascending node. We treat this angle as a close approximation of the resonant angle, in which it is diagnostic of the nature of a resonance event. This angle tends to librate about $180^\circ$ when a planet is captured in resonance in the vicinity of Cassini State 2, otherwise it usually circulates \citep{Deitrick_2018a, Quarles_2020}. Here we show that the resonant angle circulates while the planet approaches and moves away from the resonance center while an inflection point appears at the time of the crossing. This behavior reinforces that the planet was never captured into the resonance at any point in time and indeed crossed through its hyperbolic point.

\subsection{Spin-orbit Resonance Capture}

\subsubsection{Initial Conditions and Pre-analysis} \label{sec:capture_initial}
Following the spin-orbit resonance crossing experiment described in Section \ref{sec:crossing_initial}, we now consider an alternative fate for the Earth-mass planet in which it could instead become captured into the spin-orbit resonance. Here we use the exact same initial conditions as the former experiment save one: the initial precession angle. Since the outcome is probabilistic based on the value of the precession angle upon its arrival at the resonance \citep{Henrard_1982, Saillenfest_2021a}, we consider a slew of initial precession angles but handpick the case of the Earth-mass planet with an initial precession angle of $180^\circ$. We stress here that although this serves our example, this is an atypical pathway toward resonance capture since the planet crosses the resonance ``from above" ($\alpha/\nu_1$ proceeds toward unity from greater to smaller values) \citep{Ward_2004, Hamilton_2004}. We simulate this system for 4 Gyr with the same sample rate and time step as the experiment described in Section \ref{sec:crossing_initial}.

\subsubsection{Results and Discussion} \label{sec:capture_results}
\begin{figure}[!ht]
\centering
\includegraphics[width=\textwidth]{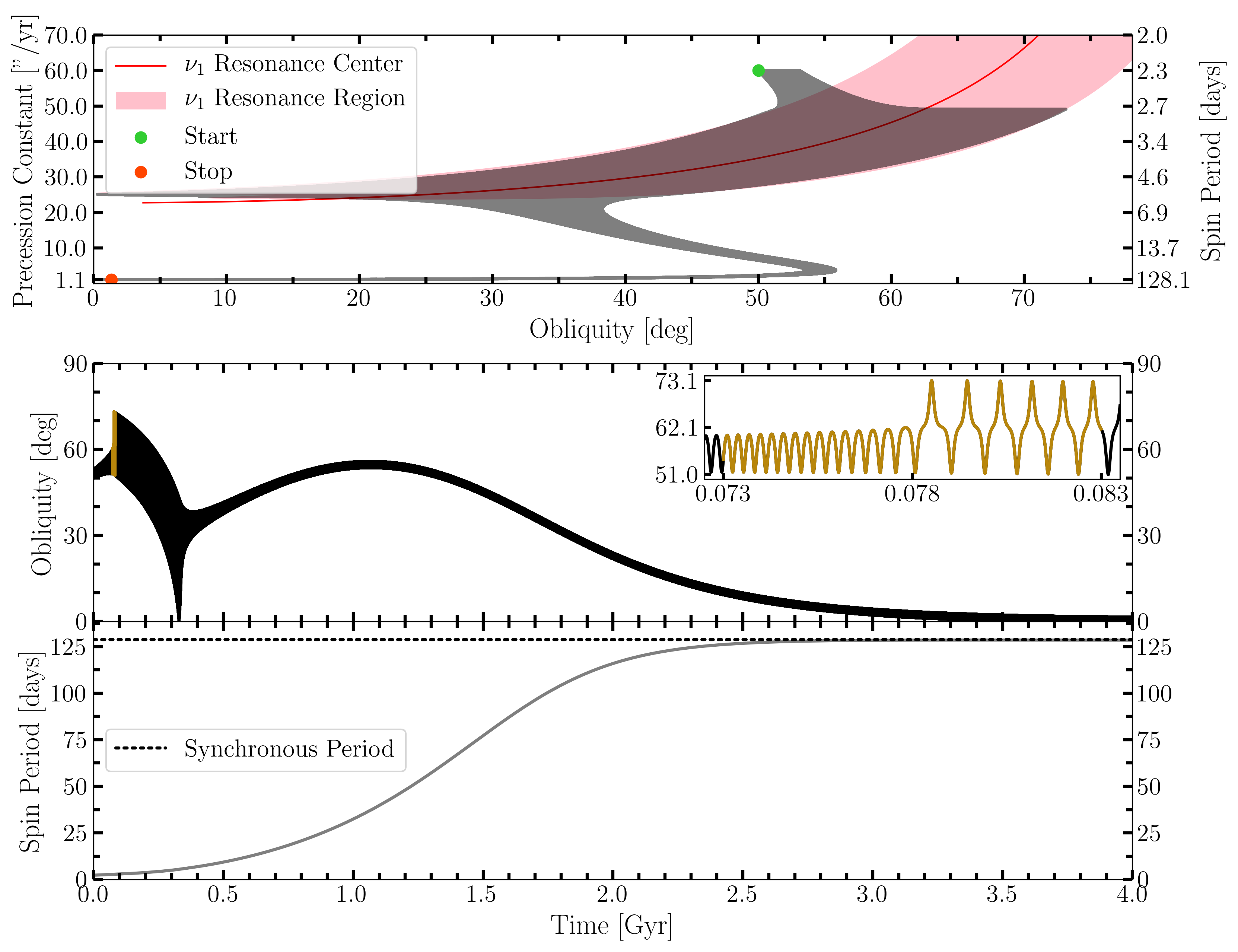}
\caption{Similar to Figure \ref{fig:crossing_plot}, here we show an alternative pathway for the evolution of the inner Earth-mass planet's precession constant, obliquity, and rotation period over the course of 4 Gyr for the Earth-Earth-Jupiter system described in Table \ref{tab:orbital}. Here the planet's initial precession angle was $180^\circ$. The middle panel includes an inset that showcases the shift in obliquity after the planet is captured into the $\nu_1$ resonance (highlighted in gold).}
\label{fig:capture_plot}
\end{figure}

\begin{figure}[!ht]
  \centering
  \includegraphics[width=0.75\textwidth]{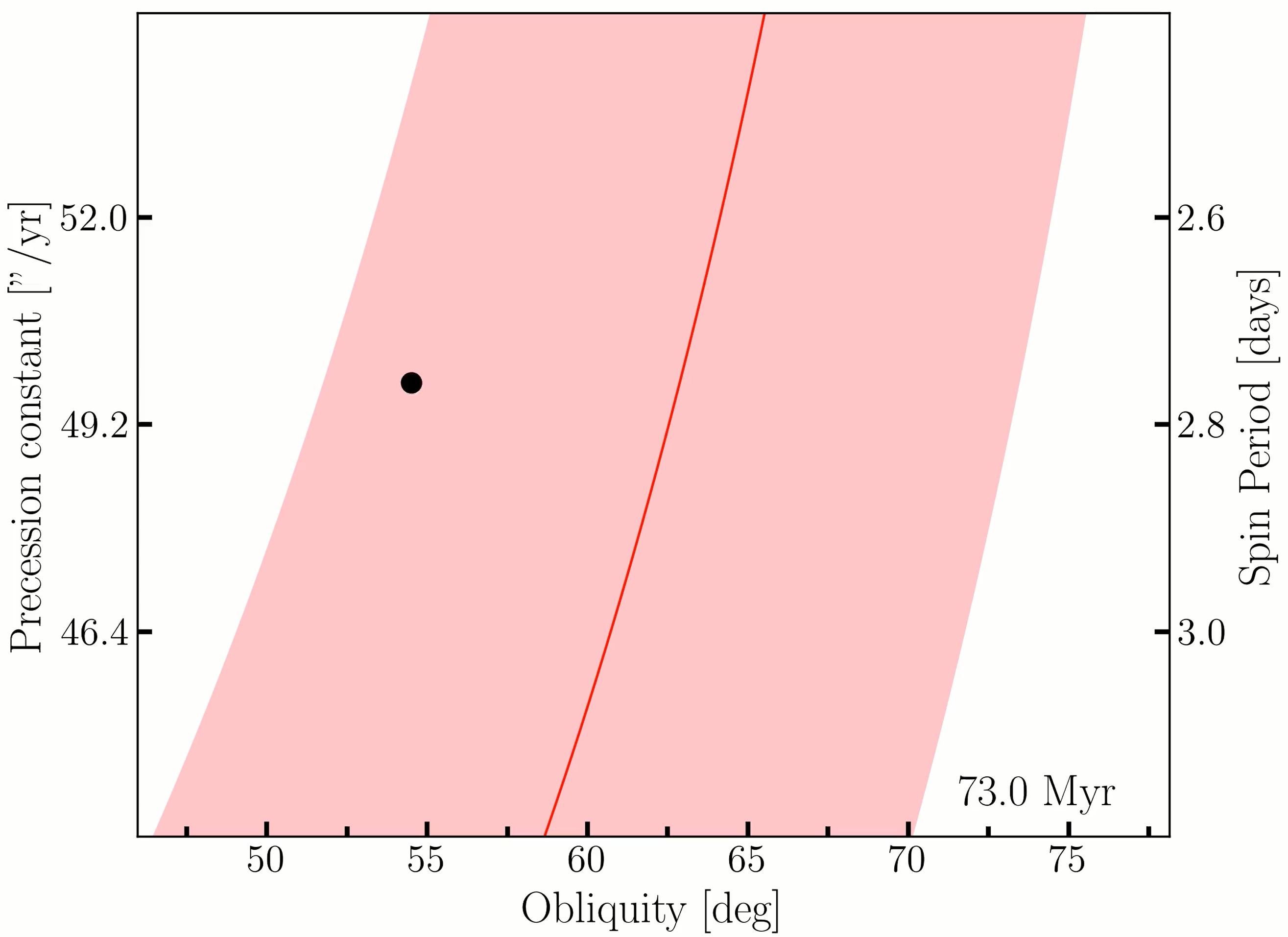}
  \caption{The real-time duration of this video is 30 seconds, during which we show a time lapse of the results from the top panel of Figure \ref{fig:capture_plot}; refer to the caption of Figure \ref{fig:crossing_plot} for a description of the chart elements. Similar to Figure \ref{fig:crossing_frame}, we represent the Earth-mass planet's parameter position with a black dot and display an associated time stamp for each frame in the lower right corner. From the range 73-83 Myr into the simulation, tidal spin torques from the star cause the Earth-mass planet's precession constant to decline while it experiences strong obliquity variations as it traverses the resonance region. The planet becomes captured upon encountering the resonance center, after which it maintains large obliquity variations about the resonance center while its precession constant continues to gradually decline. \\
  \url{https://github.com/SMKreyche/Kreyche-et-al.-2021-Videos}}
  \label{fig:capture_frame}
\end{figure}

We show the results of the spin-orbit resonance capture experiment by looking to Figure \ref{fig:capture_plot}. While the Earth-mass planet begins on a track very similar to the crossing experiment described in Section \ref{sec:crossing_results}, its spin evolution takes a very different path once it enters the resonance. After about 75 Myr into the simulation, the planet crosses the resonance center and is subsequently captured into the $\nu_1$ resonance. From here, its obliquity experiences large variations of order $\sim 20^\circ$ as it is bound to the follow the track of the resonance's center as its precession constant declines. Eventually, at around the 350 Myr mark the separatrix of the resonance disappears and the planet exits the resonance. Since the planet's ride within the resonance led to an inefficient decay of its spin rate, the planet's obliquity then trends to greater values until peaks at about $50^\circ$ and then reverses while the planet heads towards tidal equilibrium. The path that the planet's obliquity ultimately took from start to finish spanned a wide range of values with frequently large obliquity variations over short timescales. Similar to the resonance crossing experiment from Section \ref{sec:crossing_results}, this behavior would likely be detrimental toward the planet's prospects for habitability. Look to Figure \ref{fig:capture_frame} for a video that shows a time lapse of the initial resonance capture event. 

\begin{figure}[!ht]
\centering
\includegraphics[width=\textwidth]{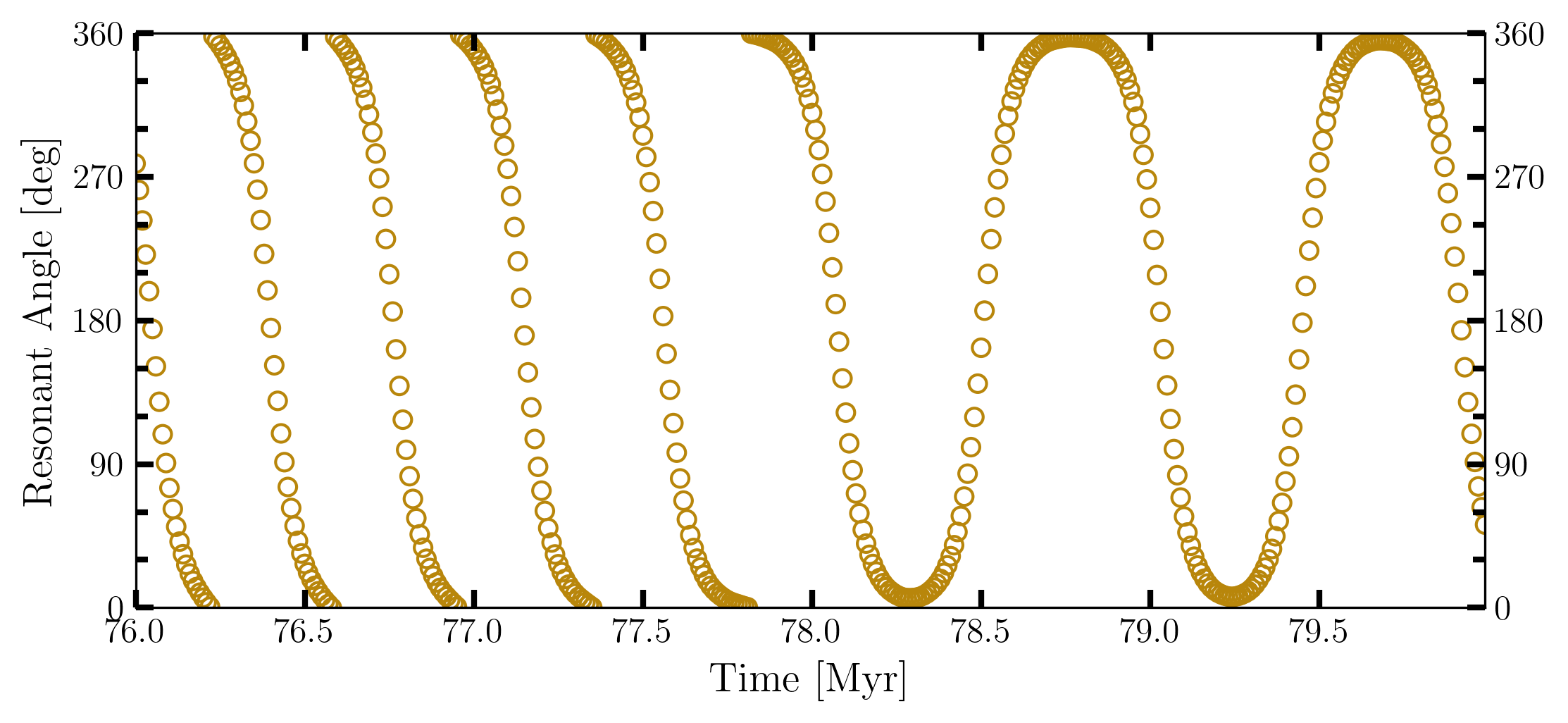}
\caption{Similar to Figure \ref{fig:crossing_analysis}, here we show an alternative evolution of the Earth-mass planet's (from the Earth-Jupiter system described in Table \ref{tab:orbital}) resonant angle over time centered about the time that the planet crossed the $\nu_1$ resonance center. In this case, the planet's initial precession angle was $180^\circ$. We see that this time, the resonant angle transitions from circulating to librating about $180^\circ$ upon the encounter, which is a strong indicator that it was captured into the resonance.}
\label{fig:capture_analysis}
\end{figure}

Why did this experiment lead to the Earth-mass planet getting captured into the resonance while the other did not? Recalling that we set this planet to have an initial precession angle of $180^\circ$ rather than the former experiment's $0^\circ$, we show the temporal evolution of its resonant angle centered around the event in Figure \ref{fig:capture_analysis}. Here we show that prior to the planet's encounter with the resonance center its resonant angle circulates. However, following the event the resonant angle then librates about $180^\circ$, albeit very widely, which is confirmation of the planet's capture into the resonance. This time, our choice of initial precession angle ensured that the conditions were ripe for capture into the resonance at the time of crossing.

\subsection{Chaotic Crossing}

\subsubsection{Initial Conditions and Pre-analysis} \label{sec:chaotic_initial}
For our final experiment, we design a scenario that exemplifies a chaotic spin-orbit resonance crossing. The resonance overlap criterion specifies that chaotic regions of parameter space exist in the case that two first-order resonances overlap. Therefore we now consider a system with an additional planet to introduce an additional forced inclination frequency \citep{Saillenfest_2019}. We present this system in Table \ref{tab:orbital} listed under system E-E-J, which shows our new system to consist of the Sun, an Earth-mass planet at 0.5 AU with an inclination of $5^\circ$, an Earth-mass planet at 0.7 AU with an inclination of $2.5^\circ$, and a Jupiter-mass planet at 3.5 AU whose orbit sits flat with an inclination of $0^\circ$. 

We conduct the same frequency analysis as described in Section \ref{sec:crossing_initial} to extract the information on the relevant modes of this system. We display this information in Table \ref{tab:freqs} listed under system E-E-J. This time, our inclusion of the additional Earth-mass planet raised the ``temperature" of the system, where the inner Earth shows two prominent driving frequencies in its inclination series. The first mode, $\nu_1$, has a frequency of $\approx -10.67$ $''$/yr and is the primary driver of the inner Earth-mass planet's orbital precession. The second mode, $\nu_2$, has a frequency of $\approx -32.25$ $''$/yr with a considerably smaller amplitude. These two frequencies have resonant widths that overlap to form a chaotic region.

We set the inner Earth's obliquity to $60^\circ$ and its precession constant to 110 $''$/yr (using Equation \ref{eqn:prec_const} with a spin period of $\approx 29.88$ hours and $J_2=6.83897 \times 10^{-4}$) so that it will begin its tidal evolution just outside of the $\nu_2$ resonance region and will tidally evolve to cross into the chaotic region. The inner Earth has the same average density, normalized polar moment of inertia, potential love number of degree 2, and constant time lag found in Table \ref{tab:physical}. We simulate this system over the course of 4 Gyr by sampling on 10,000 year intervals with a 6.46 day time step.

\subsubsection{Results and Discussion}
\begin{figure}[!ht]
\centering
\includegraphics[width=\textwidth]{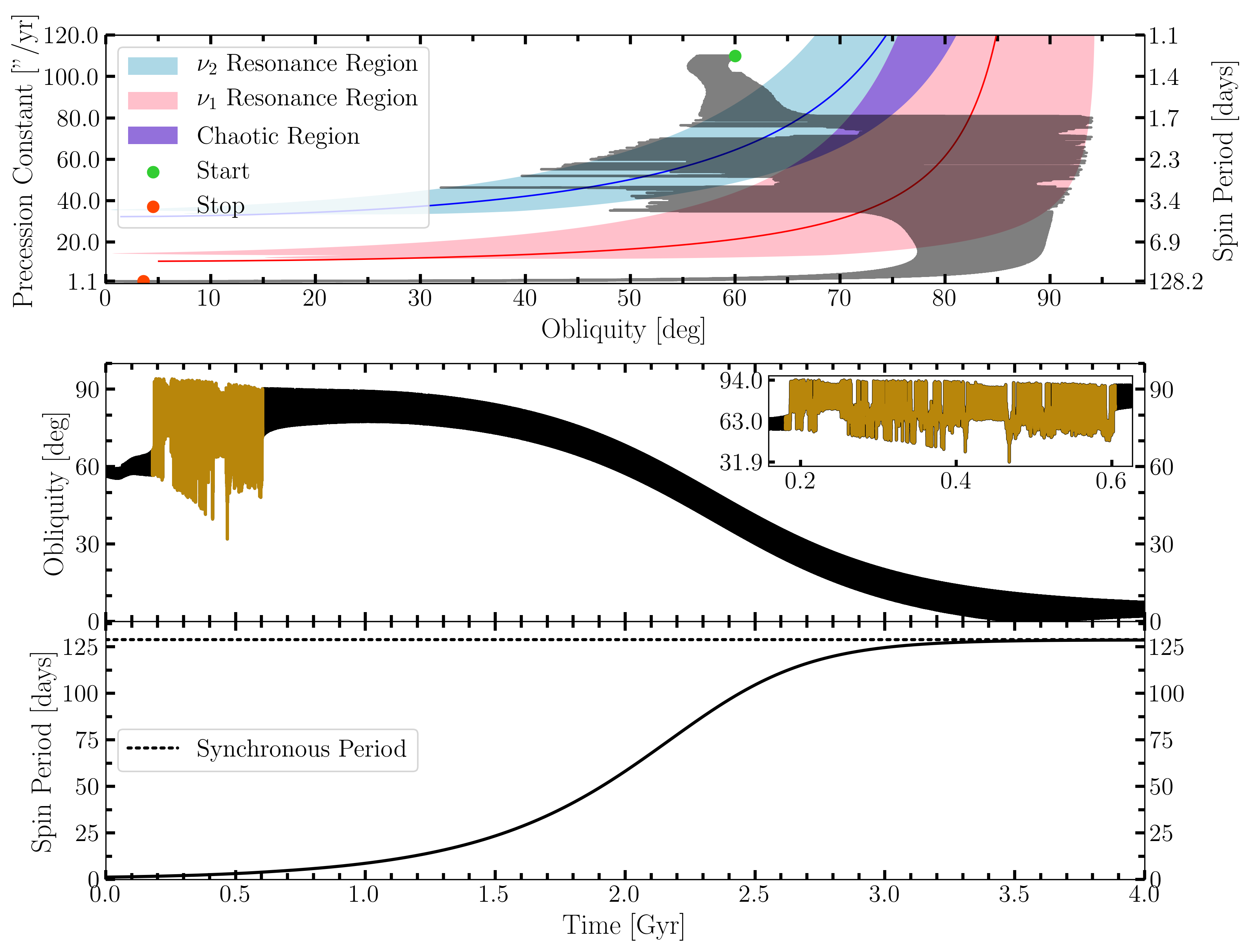}
\caption{Similar to Figure \ref{fig:crossing_plot}, here we show the evolution of the inner Earth-mass planet's precession constant, obliquity, and rotation period over the course of 4 Gyr for the Earth-Earth-Jupiter system described in Table \ref{tab:orbital}. The upper panel shows the region and center of the $\nu_2$ resonance (light blue fill and solid blue line) and the region and center of the $\nu_1$ resonance (pink fil and solid red line). The extent of their overlap exists as a chaotic region (purple fill). The middle panel includes an inset that showcases the chaotic obliquity regime from $\sim 200-600$ Myr (highlighted in gold).}
\label{fig:crossing_chaotic_plot}
\end{figure}

\begin{figure}[!ht]
  \centering
  \includegraphics[width=0.75\textwidth]{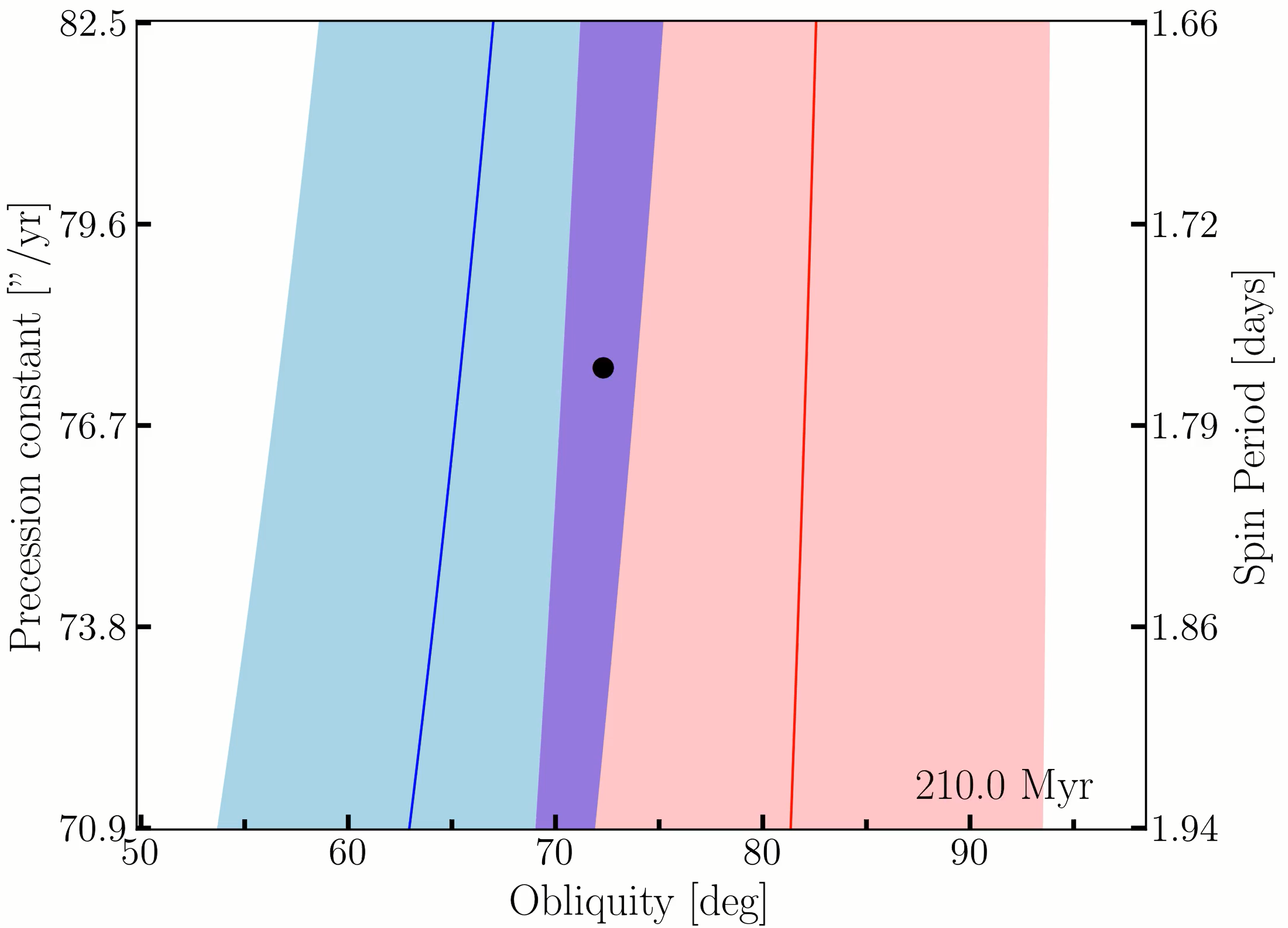}
  \caption{The real-time duration of this video is 30 seconds, during which we show a time lapse of the results from the top panel of Figure \ref{fig:crossing_chaotic_plot}; refer to the caption of Figure \ref{fig:crossing_plot} for a description of the chart elements. Similar to Figure \ref{fig:crossing_frame}, we represent the Earth-mass planet's parameter position with a black dot and display an associated time stamp for each frame in the lower right corner. From the range 210-220 Myr into the simulation, tidal spin torques from the star cause the Earth-mass planet to cross through the chaotic overlap region as its precession constant declines, causing its obliquity to undergo large and chaotic swings. \\
  \url{https://github.com/SMKreyche/Kreyche-et-al.-2021-Videos}}
  \label{fig:chaotic_crossing_frame}
\end{figure}

We show the results of the chaotic spin-orbit resonance crossing experiment in Figure \ref{fig:crossing_chaotic_plot}. These results share similarities with Figures 4a and 4b of \citet{Neron_de_Surgy_1997} that studied possible chaotic futures for the real Earth. Here the inner Earth begins with an obliquity of $60^\circ$ and experiences variations of order $5-10^\circ$ which are on par with benign variations resulting primarily from the precessional motion of its orbit. As tidal spin torques dampen the planet's precession constant and push its obliquity towards higher values, the planet crosses into the $\nu_2$ resonance after about 150 Myr. The resonance excites larger obliquity variations as the planet first crosses the center of the $\nu_2$ resonance and then into the chaotic overlap region of the $\nu_2$ and $\nu_1$ resonance. Chaos ensues, in which the planet's obliquity violently swings as much as $\sim 60^\circ$ in an unpredictable fashion within the range $30-90^\circ$. This behavior spans from 200-600 Myr into the simulation until the planet finally transitions to crossing through just the $\nu_1$ resonance and eventually exits to proceed towards tidal equilibrium where its obliquity trends towards zero and its spin period approaches the synchronous period $\sim 129$ days. Although this planetary system is fictional, its example of chaotic spin evolution implicates the probable catastrophic effects that this behavior could have on the planet's well-being. Refer to Figure \ref{fig:chaotic_crossing_frame} for a video that shows a time lapse of for a period of this chaotic behavior.

\section{Conclusion} \label{sec:conclusion}
In this article we presented our new code, \texttt{SMERCURY-T}, a numerical N-body integrator that is capable of computing the orbital and rotational evolution of a planet under the influence of tidal spin torques. Additionally, we included a module that adds the effects of the general relativistic force to planetary orbital evolution. We verified the proper implementation of these inclusions and showed the potential of \texttt{SMERCURY-T} as a powerful and versatile tool that is capable of tackling a variety of problems.

We used \texttt{SMERCURY-T} to perform a series of experiments to investigate different phenomena that a planet could experience due to the effects of tidal spin torques. These processes are likely important to the long-term dynamics of many planets. The spin-orbit resonance crossing and capture experiments demonstrated the possibility that a planet can take vastly different pathways which are determined purely by its initial conditions. While the resonance crossing experiment showed an example in which tidal spin torques drew a planet to encounter a spin-orbit resonance followed by an abrupt kick to its obliquity, the resonance capture experiment instead resulted in the planet engaging in a stable libration within the resonance while it was transported toward lower obliquity for a period of time. We also showed the results of the crossing of a chaotic region formed by the overlap of two spin-orbit resonances. The large chaotic swings of the planet's obliquity are familiar and consistent with the dynamics of some of the bodies in our own Solar System.

Although the planetary systems that we considered in this article were fictional, our general study of the nature of tidal obliquity evolution demonstrates possible scenarios that may have implications for the past, present, and future dynamics of planetary obliquity evolution. Knowing that the climatic stability and consequently, the habitability of planet, is strongly related to the nature of its obliquity evolution, understanding the dynamic role that tidal spin torques can play is critical knowledge. Large and rapid obliquity variations like those that we observed in our experiments would likely be detrimental toward a planet's climatic stability, and therefore its prospects for habitability. Although we did not investigate this claim in this article, an interesting avenue for future work could involve the application of a climate model to explore the potentially destructive consequences of these evolutionary pathways.    

For future work, \texttt{SMERCURY-T} can serve as a tool to further explore tidal obliquity evolution and seek answers to questions about the planets in our Solar System and beyond. The effects of tidal spin torques are especially relevant to planets that are in close proximity to their star, such our own system's inner terrestrial planets or those in compact exoplanet systems such as TRAPPIST-1. Future characterization of exoplanet obiquities will allow for the opportunity to perform more complete assessments of their dynamical nature.

\section{Acknowledgements}
We thank the anonymous referees for their insights and suggestions which helped to improve this manuscript. This work was supported by the NASA Habitable Worlds program, grant No. 80NSSC19K0312.

\appendix

\section{Scheme to Update Planetary Radius and $J_2$} \label{sec:appendix_a}
The inclusion of the tidal spin torque module that we discuss in Section \ref{sec:obl_tides_module} calls for the need to update the values of the spin-tracked planet's equatorial radius, $R$, and  zonal gravity coefficient, $J_2$, at each time step within a \texttt{SMERCURY-T} simulation (or at an interval set by the ``tidal tolerance" parameter). This is necessary because the effects of the tidal torque on the planet's equatorial bulge cause its spin rate, and therefore its equatorial radius and $J_2$, to decay over time. These parameters are critical for \texttt{SMERCURY-T} to compute the magnitude of the tidal torque following Equation \ref{eqn:tidal_torque} as well as to populate the planet's inertia tensor for the N-body torque calculations following \citet{Lissauer_2012}.

Sticking with the rigid body considerations described in Section \ref{sec:old}, we begin an expression for the planet's equatorial radius (like that of Equation \ref{eqn:R}) as

\begin{equation} \label{eqn:R_appendix}
    R^3 = \frac{G m_p D}{\omega_p^2} \left (\frac{2-\sqrt{4-\frac{30\omega_p^2}{G D \pi \rho}}}{10} \right) 
\end{equation}

\noindent where $m_p$, $\omega_p$, and $\rho$ are the planet's mass, spin rate, and average density, respectively. Here $G$ is the universal gravitational constant and

\begin{equation}
    D = \frac{25}{4} \left (\frac{3}{2}\overline{C}-1 \right)^2 + 1
\end{equation}

\noindent where $\overline{C}$ is the planet's normalized polar moment of inertia. Now, we can shuffle the equation $S = I_p \omega_p$ for the planet's rotational angular momentum to get

\begin{equation} \label{eqn:omega}
    \omega_p = \frac{S}{I_p} = \frac{S}{\overline{C} m_p R^2} 
\end{equation}

\noindent Here $S$ is the magnitude of the planet's spin angular momentum and $I_p$ its moment of inertia. We plug Equation \ref{eqn:omega} into Equation \ref{eqn:R_appendix} to work toward a solution for $R$. This gives

\begin{equation}
    R^3 = \frac{G {m_p}^3 D {\overline{C}}^2 R^4}{S^2} \left (\frac{2-\sqrt{4-\frac{30 S^2}{G {m_p}^2 D {\overline{C}}^2 \pi \rho R^4}}}{10} \right) 
\end{equation}

\noindent Shifting some things around, squaring both sides, and isolating the $R$ terms to one side yields:

\begin{equation}
    - \frac{40 S^2}{G {m_p}^3 D {\overline{C}}^2}R^3 + \frac{100 S^4}{G^2 {m_p}^6 D^2 {\overline{C}}^4} R^2 + \frac{30 S^2}{G {m_p}^2 D {\overline{C}}^2 \pi \rho} = 0
\end{equation}

\noindent This reduces to

\begin{equation} \label{eqn:R_solve}
    -\frac{4}{m_p}R^3 + \frac{10 S^2}{G {m_p}^4 D {\overline{C}}^2} R^2 + \frac{3}{\pi \rho} = 0
\end{equation}

This equation takes the form $ax^3 + bx^2 + cx + d = 0$, for which we solve for $R$ with the use of Cardano's general cubic formula \citep{Cardano_1993}. We obtain the corresponding value of $\omega_p$ by plugging the solution for $R$ back into Equation \ref{eqn:omega}. Finally, we calculate the planet's $J_2$ with Equation \ref{eqn:J_2}.

















\bibliography{Kreyche_2021}

\end{document}